\documentclass[aps,journal=prb,twocolumn,noeprint]{revtex4-2}
\usepackage{graphicx}
\usepackage{epstopdf}
\usepackage{bmpsize}
\usepackage{braket}
\usepackage{siunitx}
\usepackage{tabularx}
\usepackage{amsmath,amsfonts,amsthm,bm}
\usepackage{bigints}
\usepackage{xcolor}
\usepackage{tikz}
\usetikzlibrary{decorations.pathreplacing}
\usepackage{placeins}
\usepackage{colortbl}
\usepackage{dsfont}

\usepackage{ragged2e}

\usepackage[caption=false]{subfig}

\newcommand{\phantomsubfloat}[1]{
    {
        \captionsetup[subfigure]{labelformat=empty}
        \subfloat[][]{#1}
    }%
}

\newcommand{\ph}{\phantom{\dagger}}
\newcommand{\pd}{\phantom{\dagger}}

\usepackage{comment} 

\usepackage[bookmarks=false]{hyperref}
\hypersetup{colorlinks=true, citecolor=blue, urlcolor=blue, linkcolor=blue}

\usepackage{mathtools}

\newcommand\vertarrowbox[3][6ex]{%
  \begin{array}[t]{@{}c@{}} #2 \\
  \left\uparrow\vcenter{\hrule height #1}\right.\kern-\nulldelimiterspace\\
  \makebox[0pt]{\scriptsize#3}
  \end{array}%
}

\makeatletter
\renewcommand{\maketag@@@}[1]{\hbox{\m@th\normalsize\normalfont#1}}%
\makeatother

\def\!{\mskip-\thinmuskip}

\newcommand{\figref}[1]{(\ref{#1})}
\graphicspath{{plots/}}

\begin{document}
\title{Vibrational Instabilities in Charge Transport through Molecular Nanojunctions: The Role of Nonconservative Current-Induced Electronic Forces}
\author{Martin Mäck}
\email[Corresponding author: Martin Mäck \\ Email:
]{martin.maeck@physik.uni-freiburg.de}
\affiliation{Institute of Physics, University of Freiburg, Hermann-Herder-Strasse 3, 79104 Freiburg, Germany}
\author{Riley J. Preston}
\affiliation{Institute of Physics, University of Freiburg, Hermann-Herder-Strasse 3, 79104 Freiburg, Germany}
\author{Michael Thoss}
\affiliation{Institute of Physics, University of Freiburg, Hermann-Herder-Strasse 3, 79104 Freiburg, Germany}
\author{Samuel L. Rudge}
\affiliation{Institute of Physics, University of Freiburg, Hermann-Herder-Strasse 3, 79104 Freiburg, Germany}

\begin{abstract}
\noindent Understanding the current-induced vibrational dynamics in molecular nanojunctions is critical for gaining insight into the stability of such systems. While it is well known that Joule heating at higher bias voltages plays an important role for the stability of the nanojunction, a different mechanism caused by current-induced nonconservative forces has been reported to cause vibrational instabilities already at much lower voltages. In this paper, we apply a mixed quantum-classical approach based on electronic friction and Langevin dynamics to a model system for which vibrational instabilities have previously been reported. The electronic friction and other electronic forces can be calculated exactly using nonequilibrium Green's functions. We perform Langevin simulations of the vibrational dynamics to calculate the steady-state vibrational excitation and compare the results to a simplified eigenmode analysis, identifying the parameter regimes in which runaway modes appear.
\end{abstract}

\maketitle

\section{Introduction}\label{sec:Introduction}

\noindent Molecular nanojunctions are devices built by attaching single molecules to two or more metallic leads and have been considered as potential candidates to replace silicon-based devices, allowing for further miniaturization of electronic circuits \cite{vonHippel1956,Aviram1974,Nitzan2001,Nitzan2003,Cuevas2010, Zimbovskaya2011, Bergfield2013, Thoss2018}. Moreover, molecular nanojunctions can be used to investigate novel transport phenomena stemming from the quantum mechanical nature of such systems \cite{Galperin2007,Koch2006b, vonOppen2008, Hrtle2011, Persson1997, Ring2020, Sabater2015,PhysRevB.105.195435}. 

An important concern regarding the operating ability of molecular nanojunctions is their stability under nonequilibrium conditions at finite bias voltage. This is largely determined by the interaction between the vibrational and electronic degrees of freedom (DOFs) within the molecule. It is well known that this electronic-vibrational interaction can lead to current-induced or Joule heating and, at sufficiently high bias voltages, to vibrational instabilities \cite{Segal2002,Huang2007,Hrtle2008,C1CP21161G,Tao2006,doi:10.1126/science.1146556,Ioffe2008,Montgomery2002,Pecchia2007,Schinabeck2018} and current-induced bond rupture \cite{Erpenbeck2018,Erpenbeck2020,Ke2021, Ke2023,Li2015,Li2016,Capozzi2016,doi:10.1126/science.272.5260.385,doi:10.1021/nl801669e,PhysRevB.78.045434}. Beyond current-induced heating, recent investigations have also highlighted a different mechanism causing vibrational instabilities in nanojunctions that arises from nonconservative current-induced forces \cite{L2010, L2011}. This proposed mechanism is based on a classical treatment of the vibrational DOFs, where the influence of the quantum electronic degrees of freedom appears via current-induced forces, and should be influential when multiple vibrational modes are involved in transport. Essentially, the highly nonconservative landscape of electronic forces in out-of-equilibrium scenarios \cite{Dundas2009, SORBELLO1998159,PhysRevLett.114.096801,Gunst2013,Christensen2016,Lue2019} can induce ``runaway" vibrational eigenmodes at much lower bias voltages than that required for instabilities caused by simple Joule heating, especially when the vibrational dynamics are pulled into elliptical trajectories via the nonadiabatic Berry force \cite{L2010, L2011,L2012,PhysRevLett.107.036804,Todorov2014}. 

Understanding the connection between the vibronic dynamics of molecular nanojunctions and corresponding vibrational instabilities is evidently crucial in creating mechanically stable molecular-based devices. However, simulating the full nonequilibrium quantum mechanical dynamics of molecular nanojunctions is a difficult task, as one must include not only the interaction between electronic and vibrational DOFs within the molecule, but also the interaction of the molecule with the continuum of electronic states in the leads. Although there exist fully quantum mechanical methods for such systems, they often rely on a perturbative treatment of either the molecule-lead coupling, as in the case of Born-Markov master equations \cite{Koch2006a,Leijnse2008,Koch2006b,Koch2005,Rudge2019a,Rudge2019b,PhysRevB.79.205303}, or the electronic-vibrational interaction, such as in nonequilibrium Green's function (NEGF) theory \cite{Mitra2004,Erpenbeck2015,Galperin2006,Ryndyk2006,Hrtle2008,Novotn2011}. There exist also approaches that treat the vibrational DOFs quantum mechanically and numerically exactly, such as the hierarchical equations of motion (HEOM) approach \cite{Tanimura1989, Schinabeck2020,Jin2008, Hrtle2013,Hrtle2018,Schinabeck2016} and the multilayer multiconfigurational time-dependent Hartree
method \cite{10.1063/1.3173823, 10.1063/1.3660206,10.1063/1.4965712}. However, they quickly become prohibitively numerically costly in regimes requiring a large number of vibrational basis states.

To overcome this limitation, one can employ a mixed quantum-classical approach, in which the vibrational DOFs within the molecule are treated classically while being influenced by quantum mechanical electronic DOFs, both in the molecule and the leads. A widely used mixed quantum-classical method is based on electronic friction and Langevin dynamics \cite{HeadGordon1995,Bode2012,L2012, Dou2016, Dou2016_2,Maurer2016, Dou2017, Dou2017_1, Dou2018, Chen2018, Chen2019, Preston2021, PhysRevB.83.115420,10.1063/5.0019178}.
For systems without strong intrasystem interactions, NEGFs are an accurate and numerically efficient method for calculating the electronic friction \cite{Dou2017,Dou2016, Dou2016_2, Preston2020,Preston2021,Preston2022} . Moreover, recently, the theory of electronic friction has also been extended to the HEOM method, which allows for the description of strong intrasystem interactions \cite{Rudge2023,Rudge2024,10.1063/5.0222076}.

In addition to numerical efficiency, describing the vibrational dynamics using Langevin dynamics can offer valuable insight into the electronic forces acting on the molecular vibrations \cite{Teh2021, Teh2022}, which otherwise can be challenging to extract from a fully quantum mechanical description of the system. This is highly relevant to this work, as the vibrational instabilities arising specifically from nonconservative electronic forces have been identified from and explored solely within the Langevin dynamics picture \cite{L2010, L2011,L2012,Christensen2016,Lue2019}. Furthermore, these previous investigations relied on additional approximations, calculating eigenmode solutions to the Langevin equation by neglecting the stochastic force and evaluating the current-induced forces at the Fermi energy only. These have been successful in identifying the possibility of new mechanisms for vibrational instability, with even some potential experimental evidence \cite{Sabater2015}, but a key remaining question is whether these instabilities appear in the full dynamics obtained by solving the Langevin equation directly. 

With this motivation in mind, in this work, we apply the combined NEGF and Langevin dynamics (NEGF-LD) approach to a two-level, two-mode system for which vibrational instabilities caused by nonconservative current-induced forces have previously been reported ~\cite{L2012}. We perform Langevin simulations of the vibrational dynamics to calculate the vibrational excitation in both modes, which we then use to characterize the junction stability. Contrary to previous reports, these simulations show that instabilities only occur for degenerate vibrational modes and that the Berry force plays only a minor role in the vibrational dynamics. Additionally, we perform an eigenmode analysis motivated by the work from Refs.~\cite{L2010, L2011,L2012} and demonstrate under which conditions such reduced pictures are reliable predictors of instability.

The paper is structured as follows. In Sec.~\ref{sec: Model}, we introduce a general model of a molecule interacting with metal leads. In Sec.~\ref{sec: Nonequilibrium Transport Theory}, we give a short overview of the NEGF approach and show how to describe the vibrational dynamics using a Langevin type equation and the calculation of the corresponding electronic forces using NEGFs. In Sec.~\ref{sec: Results}, we apply the approach to a model system for which vibrational instabilities at low bias voltages have previously been reported. 

Since we have to differentiate between classically and quantum mechanically treated coordinates and momenta, we will explicitly denote the vectors of position and momentum operators as $\hat{\bm{x}}$ and $\hat{\mathbf{p}}$, while their classical counterparts will be denoted by $\bm{x}$ and $\mathbf{p}$. Moreover, in this work, we use units where $e=\hbar=1$.  

\section{Model} \label{sec: Model}

In this section, we introduce the general model of a molecular nanojunction considered within this work. The total Hamiltonian of the setup is 
\begin{align}
    H = \: & H_{\text{mol}} + H_{\text{leads}} + H_{\text{mol-leads}},
\end{align}
where $H_{\text{mol}}$ is the molecular Hamiltonian, $H_{\text{leads}}$ is the Hamiltonian of the leads, and $H_{\text{mol-leads}}$ describes the interaction between the two.

We are specifically interested in a vibronic model containing two harmonic vibrational modes linearly coupled to two electronic degrees of freedom without electron-electron interactions. The corresponding molecular Hamiltonian takes the form,
\begin{align}
    H_\text{mol} = \: & \underbrace{\sum_{mn} h_{mn}(\hat{\bm{x}})d^\dagger_{m} d^{\pd}_{n}}_{= H^{\text{el}}_\text{mol}} + \underbrace{\sum_{i}\frac{\omega_{i}}{2}\left(\hat{x}_{i}^{2} + \hat{p}_{i}^{2}\right)}_{= H^{\text{vib}}_\text{mol}}, \label{eq: molecular Hamiltonian}
\end{align}
which has been written in mass- and frequency-scaled dimensionless vibrational coordinates. Here, $H^{\text{vib}}_\text{mol}$ contains not only the vibrational kinetic energy, but also the potential energy of the vibrations without electronic influence, while $H^{\text{el}}_\text{mol}$ contains the electronic energy and the electronic-vibrational interaction. 

The $i$th vibrational DOF has frequency $\omega_{i}$, coordinate $\hat{x}_{i}$, and momentum $\hat{p}_{i}$, which are collected into the vectors $\hat{\bm{x}} = \left(\hat{x}_{1},\hat{x}_{2}\right)$ and $\hat{\mathbf{p}} = \left(\hat{p}_{1},\hat{p}_{2}\right)$, respectively. The energies of the electronic degrees of freedom, are contained in the matrix of single-particle energies and interactions, $h(\hat{\bm{x}})$, which in the single-particle basis and for this work has the form 
\begin{align}
    h(\hat{\bm{x}}) = \: & \left(\begin{array}{c c}
       \varepsilon_{0} + \lambda_{2} \hat{x}_{2} & t + \lambda_{1} \hat{x}_{1} \\
       t + \lambda_{1} \hat{x}_{1} & -\left(\varepsilon_{0} + \lambda_{2} \hat{x}_{2}\right)
    \end{array}\right).
\end{align}
Here, the operators $d^{\dag}_{m}$ and $d^{\pd}_{m}$ create and annihilate an electron at the $m$th energy level with energy $\varepsilon_{m}(\hat{\bm{x}}) = h_{mm}(x_{2})$, respectively. The off-diagonal terms, $h_{m,n\neq m}(\hat{\bm{x}})$, describe hopping between the $m$th and $n$th levels, and depend on some direct hopping component with strength $t$ as well as vibrationally assisted hopping, $\lambda_{1}\hat{x}_{1}$. 

The left $(L)$ and right $(R)$ leads are modeled as reservoirs of noninteracting electrons, 
\begin{equation}
    H_{\text{leads}} =  \sum_{\alpha \in \{L,R\}} \sum_{k } \varepsilon^{\ph}_{k \alpha} c^{\dagger}_{k \alpha}c^{\ph}_{k \alpha}.
\end{equation} 
Here, the energy of state $k$ in lead $\alpha$ is given by $\varepsilon_{k \alpha}$, while the corresponding creation and annihilation operators are $c^{\dagger}_{k \alpha} $ and $c_{k \alpha}$, respectively. The leads are held at local equilibrium, such that they have a well-defined chemical potential, $\mu_{\alpha}$, and temperature, $T$. We apply nonequilibrium conditions via a bias voltage across the junction, $\Phi = \mu_{L} - \mu_{R}$, which is applied symmetrically, such that $\mu_{L} = -\mu_{R} = e\Phi/2$.

The interaction between the molecule and the leads is given by,
\begin{align}
    H_{\text{mol-leads}} = \: & \sum_{k,\alpha}\sum_{m} V_{k\alpha,m}\left(c_{k {\alpha}}^{\dagger}d^{\ph}_m +d_{m}^{\dagger}c^{\ph}_{k {\alpha}}\right),
\end{align}
where $V_{k\alpha,m}$ describes the coupling strength between state $m$ in the molecule and state $k$ in lead $\alpha$. For noninteracting reservoirs and a linear molecule-lead coupling, the influence of the leads on the molecular dynamics is completely described by two-time correlation functions, which are in turn characterized in terms of the spectral density of each lead: 
\begin{align}
    \Gamma_{\alpha,m m'} (\epsilon) = \: & 2 \pi \sum_{k} V^{\ph}_{k\alpha,m}V^{\ph}_{k\alpha,m'} \delta(\epsilon - \varepsilon_{k \alpha}).
\end{align}
To connect with previous investigations of runaway modes in charge transport through molecular nanojunctions, we will exclusively work in the wideband limit, such that the spectral density is a constant: $\Gamma_{\alpha,m m'} (\epsilon) = \Gamma_{\alpha,m m'}$. Additionally, we couple only level $1$ to the left lead and level $2$ to the right lead, such that $\Gamma_{L,22} = \Gamma_{R,11} = \Gamma_{\alpha,21} = \Gamma_{\alpha,12} = 0$, and further assume that the remaining coupling strengths are the same: $\Gamma_{L,11} = \Gamma_{R,22} = \Gamma$. Henceforth, we will refer to $\Gamma$ as the molecule-lead coupling strength.

\section{Nonequilibrium Transport Theory} \label{sec: Nonequilibrium Transport Theory}

\noindent In this section, we introduce and discuss the approaches we use to investigate vibrational instabilities in nonequilibrium charge transport through molecular nanojunctions. First, in Sec.~\ref{subsec: Electronic Friction and Langevin Dynamics}, we briefly outline how one obtains a mixed quantum-classical Langevin equation from the full quantum dynamics, and give expressions for the resulting electronic forces in terms of nonequilibrium Green's functions. Then, in Sec.~\ref{subsec: Langevin simulations and Observables} we discuss the numerical details of solving the Langevin equation and how one obtains expectation values of observables in the steady-state. Finally, in Sec.~\ref{subsec: Dynamics from an Eigenmode Analysis}, we introduce how one can analyze the Langevin equation via an approximate eigenmode picture.  

\subsection{NEGF and Langevin Dynamics (NEGF-LD)} \label{subsec: Electronic Friction and Langevin Dynamics}

For a comprehensive review of how a mixed quantum-classical Langevin equation for the vibrational degrees of freedom is obtained, we refer the interested reader to Refs.~\cite{L2012,Chen2018,Chen2019,Rudge2023}. Here, we will briefly outline the approach, arriving at the final equation of motion and expressions for the electronic forces within the framework of Keldysh NEGFs. 

To start, without any approximations and under a completely quantum treatment, the time evolution of the vibrational degrees of freedom can be written via the Feynman-Vernon influence functional, which incorporates the effect of the electronic degrees of freedom in the molecule and leads via an effective action. As shown in Refs.~\cite{L2012,Chen2018}, a classical equation of motion for the vibrations is obtained by moving to Wigner coordinates and expanding to second-order in the quantum difference to the classical path, 
\begin{align}
\ddot{x}_i = \: & -\omega_{i}^{2}x_{i} \underbrace{-\omega_{i}\text{Tr}_{\text{el}}\left\{\frac{\partial H_\text{mol}(\bm{x}(t))}{\partial x_i}\rho_{\text{el}}[\bm{x}(t)]\right\}}_{=F_{i}[\bm{x}(t)]} + f_i(t), \label{eq: exact_classical_force}
\end{align}
where $x_{i}$ represents the classical or average path of the quantum vibrational coordinate $\hat{x}_{i}$, and where we have explicitly used the mass- and frequency-scaled units from Eq.\eqref{eq: molecular Hamiltonian}. Here, $F_{i}[\bm{x}(t)]$ contains the average electronic force as well as the force due to the vibrational potential, and $f_{i}(t)$ is a Gaussian stochastic force describing fluctuations away from the average force. The electronic density matrix at time $t$, $\rho_{\text{el}}[\bm{x}(t)]$, is a functional of the vibrational trajectory. Note that one can also obtain this equation by starting with the Liouville-von Neumann equation for the total density matrix and performing a second-order approximation to the partial Wigner transform with respect to the vibrational degrees of freedom \cite{Rudge2023,Dou2018}.

For the model introduced in Sec.~\ref{sec: Model}, one can write the average electronic force in terms of Keldysh NEGFs\cite{Bode2012, Preston2020, Dou2017},
\begin{align}
F_{i}[\bm{x}(t)] = \: & i  \text{Tr}\left\{ \frac{\partial h(\bm{x})}{\partial x_i} G^<(t,t)\right\},
\end{align}
where the $\text{Tr}(\dots)$ indicates a trace over the single-particle Hilbert space upon which $h(\bm{x})$ acts, and the lesser Green's function is defined in the Heisenberg picture as
\begin{equation}
G_{mn}^<(t,t') = i\text{Tr}_\text{el} \left\{d_n^\dag (t') d_m (t) \rho_\text{el}(t_0) \right\}.
\end{equation}
The density matrix for the electronic subsystem, $\rho_{\text{el}}$, is assumed to be factorized at time $t_0$ as
 \begin{align}
     \rho_{\text{el}}(t_0) & = \rho_{\text{mol,el}}(t_0)\rho_{\text{leads,el}}(t_0).
 \end{align}
 Here, $\rho_{\text{mol,el}}(t_0)$ is the initial electronic state of the molecule and $\rho_{\text{leads,el}}(t_0)$ is the initial state of the leads, which is a tensor product of Gibbs' states:
 \begin{equation} \label{gibbs}
     \rho_{\text{leads,el}}(t_0)= \prod_{\alpha} \frac{e^{-\left(H_{\text{leads},\alpha} - \mu_{\alpha}\right)/k_{\text{B}}T}}{\text{Tr}_{\text{leads},\alpha}\left[ e^{-\left(H_{\text{leads},\alpha} -\mu_{\alpha} \right)/{k_{\text{B}}T}}\right]}.
\end{equation}
In the Keldysh formalism, we take $t_0 \rightarrow -\infty$, such that correlations between the molecule and leads are accounted for explicitly within the theory \cite{Altland_Simons_2023}, before numerical simulations are initiated at $t=0$.

The correlation function of the stochastic force is similarly expressed in terms of NEGFs as \cite{Preston2022, Dou2017_1}
\begin{align}
D_{ij}(t,t') &= \text{Tr}_{\text{el}}\left\{ f_i (t) f_{j}(t')\rho_\text{el}(t_0)\right\}
\\
&= \text{Tr}\left\{  \frac{\partial h(\bm{x})}{\partial x_i} G^> (t,t') \frac{\partial h(\bm{x})}{\partial x_j} G^<(t',t)\right\},
\end{align}
where the greater Green's function is defined as
\begin{equation}
G_{mn}^>(t,t') = -i\text{Tr}_\text{el} \left\{ d_m (t) d^\dag_n (t')\rho_\text{el}(t_0) \right\}.
\end{equation}

Although we have closed expressions for the electronic forces, at this point the dynamical problem is still quite difficult to solve, since the electronic degrees of freedom rely explicitly on the history of the vibrational trajectory, as does the correlation function of the stochastic force. However, if one takes the limit of weak nonadiabaticity in the form of a near-adiabatic approximation,  $F_{i}[\bm{x}(t)]$ can be expanded in powers of the momenta and Eq.\eqref{eq: exact_classical_force} takes the form of a Langevin equation \cite{10.1063/5.0153000, Preston2022, Dou2018}:
\begin{align}
    \ddot{x}_{i} = \: & -\omega_{i}^{2}x_{i} + F^{\text{ad}}_i(\bm{x}) - \sum_{j} \gamma_{ij}(\bm{x}) \dot{x}_{j} +  f_i(t), \label{langevin_equation}
\end{align}
where $f_i(t)$ is now a Gaussian random force with white noise,
\begin{align}
    D_{ij}(t,t') = \: & D_{ij}(\bm{x}) \delta(t-t').
\end{align}
The second term on the righthand side of Eq.\eqref{langevin_equation} is the average adiabatic electronic force, which is calculated for a frozen vibrational frame $\boldsymbol{x}$. It can be written in the language of Keldysh NEGFs as
\begin{equation}
    F^{\text{ad}}_i(\bm{x}) = i\int^\infty_{-\infty} \frac{d\epsilon}{2\pi} \text{Tr} \left\{ \frac{\partial h (\bm{x})}{\partial x_i} \tilde G_{(0)}^<(\epsilon, \bm{x})\right\},
    \end{equation}
    where $\tilde G_{(0)}^{</>}$ is the adiabatic lesser/greater Green's function in the energy domain, defined as
\begin{equation}
\tilde G_{(0)}^{</>} = \tilde G_{(0)}^R \tilde \Sigma^{</>} \tilde G_{(0)}^A.
\end{equation}
For the sake of brevity, we have subdued the functional dependence of the Green's functions on $(\epsilon, \bm{x})$. The retarded/advanced Green's functions take the standard form,
\begin{equation}
\tilde{G}_{(0)}^{R/A} = \Big(\epsilon \hat{I}-h-\tilde{\Sigma}^{R/A}\Big)^{-1}.
\label{GRA}
\end{equation}
By taking the wideband limit for the leads, the self-energies for an individual lead, $\alpha$, take the form,
\begin{equation}
\tilde{\Sigma}^R_{\alpha} = -\frac{i}{2}\Gamma_{\alpha},
\;\;\;\;\;\;
\tilde{\Sigma}^A_{\alpha} = \frac{i}{2}\Gamma_{\alpha},
\label{sigmaAR}
\end{equation}
\begin{equation}
\tilde{\Sigma}^<_{\alpha}(\epsilon) = if_\alpha(\epsilon)\Gamma_{\alpha},
\;\;\;\;\;\;
\tilde{\Sigma}^>_{\alpha}(\epsilon) = -i[1-f_\alpha(\epsilon)]\Gamma_{\alpha},
\label{sigma_lesser}
\end{equation}
with the Fermi-Dirac function given by
\begin{equation}
    f_{\alpha}(\epsilon)= \frac{1}{1+e^{ \left(\epsilon -\mu_\alpha \right)/k_{\text{B}}T}}.
\end{equation}
The total self-energy, which is a sum over the self-energy contributions from each lead, is written without a lead subscript, $\tilde{\Sigma}^R=\sum_\alpha \tilde{\Sigma}^R_{\alpha}$.

The correlation function of the stochastic force can also be written in terms of adiabatic Green's functions \cite{Preston2022},
\begin{equation}
   D_{ij}(\bm{x}) = \int^\infty_{-\infty} \frac{d\epsilon}{2\pi} \text{Tr} \left\{ \frac{\partial h(\bm{x})}{\partial x_i} \tilde G_{(0)}^>(\epsilon, \bm{x}) \frac{\partial h(\bm{x})}{\partial x_j} \tilde G_{(0)}^<(\epsilon, \bm{x})\right\}.
\end{equation}

The electronic friction arises as a first-order nonadiabatic correction to the adiabatic electronic force, taking the form 
\begin{equation}
    \gamma_{ij}(\bm{x}) = -i\omega_{i}\int^\infty_{-\infty} \frac{d\epsilon}{2\pi} \text{Tr} \left\{ \frac{\partial h(\bm{x})}{\partial x_i} \tilde G_{(1),x_j}^<(\epsilon, \bm{x})\right\},
    \label{friction_tensor}
    \end{equation}
where $\tilde G_{(1),x_j}^<$ is the first nonadiabatic correction to the adiabatic lesser Green's function due to the motion of $x_j$, which is given by \cite{Preston2021}
\begin{align}
\tilde{G}_{(1),x_j}^{<} = \: & \frac{1}{2i}\tilde{G}_{(0)}^{R}\Bigg(\tilde \Sigma^< \tilde G^A_{(0)}\left[\tilde G^A_{(0)},\frac{\partial h}{\partial x_j}\right]_- \nonumber \\
& + \frac{\partial h}{\partial x_j}\tilde{G}_{(0)}^{R}\frac{\partial\tilde{\Sigma}^{<}}{\partial\epsilon}+\tilde{G}_{(0)}^{<}\frac{\partial h}{\partial x_j} + \text{h.c.}\Bigg)\tilde{G}_{(0)}^{A}.
  \label{G1L}
\end{align}
Note that we have also included the vibrational frequency in the definition of the friction tensor in Eq.\eqref{friction_tensor}, which originates from the dimensionless coordinates chosen for this work.

Physically, the electronic friction describes dissipation of vibrational energy via coupling to the creation of electron-hole pairs (EHPs) in the leads, while the stochastic force describes the opposite process of energy being injected into the vibrational degrees of freedom via the decay of EHPs. In equilibrium, at zero bias voltage, these two processes are balanced via the fluctuation-dissipation theorem,
\begin{align}
    D_{ij}(\bm{x}) = \: & k_{B}T \gamma_{ij}(\bm{x}).
\end{align}
Furthermore, at equilibrium, it has been shown that, for real-valued electronic Hamiltonians, $\bm{F}^{\text{ad}}(\bm{x})$ is conservative, and the electronic friction tensor is symmetric and positive-definite for all $\bm{x}$,
\begin{align}
    \sum_{ij} \dot{x}_{i}\gamma_{ij}(\bm{x})\dot{x}_{j} \geq 0,
\end{align}
indicating that it has an overall damping effect on the vibrational dynamics. 

In contrast, none of these properties are guaranteed out of equilibrium at a finite bias voltage. For example, in the electronic forces picture, Joule heating manifests itself as a breaking of the fluctuation-dissipation theorem, where the correlation function of the stochastic force is larger than the electronic friction at vibrational coordinates for which the electronic energies lie within the bias window. Essentially, the stochastic force describes the transfer of energy to the vibrational modes not only from the decay of electron-hole pair excitations around the chemical potentials, but also from electrical current through the junction \cite{Rudge2023,Rudge2024}. Additionally, 
$\bm{F}^{\text{ad}}(\bm{x})$  generally has a nonconservative component in nonequilibrium, and the electronic friction tensor is not necessarily symmetric or positive definite. 

Consequently, at finite bias voltage, we can split the friction tensor into a symmetric and antisymmetric part 
\begin{align}\label{eq:friction_sym_antisym}
    \boldsymbol{\gamma} = \: & \underbrace{\frac{\boldsymbol{\gamma} + \boldsymbol{\gamma}^{T}}{2}}_{\boldsymbol{\gamma}^{\text{S}}} + \underbrace{\frac{\boldsymbol{\gamma} - \boldsymbol{\gamma}^{T}}{2}}_{\boldsymbol{\gamma}^{\text{A}}}.
\end{align}
The antisymmetric part of the friction tensor describes a Lorentz-type force arising from the Berry curvature, which, although it does no work, pulls the vibrational dynamics into curved trajectories. 

\subsection{Langevin Simulations and Observables} \label{subsec: Langevin simulations and Observables}

We simulate the vibrational dynamics, by solving the Langevin equation in Eq.\eqref{langevin_equation}. In this work, we exclusively use the ABOBA algorithm, which belongs to a stable and efficient family of approaches for solving stochastic differential equations, in which the generator of the Langevin equation is split via a Trotter decomposition into three parts (A,B,O) and applied in a specific order. For a more comprehensive outline of the method, see Ref.~\cite{e19120647} and also the summary in Ref.~\cite{Rudge2024} as it pertains to charge transport in molecular nanojunctions.

To calculate expectation of observables, we need to average over many trajectories, $N_{\text{traj}}$, generated via the Langevin equation. We assume that the molecule is initially in its vibrational ground state, such that we can sample $N_{\text{traj}}$ initial conditions from the Wigner distribution of the ground state of a two-dimensional harmonic oscillator,
\begin{equation}
    \rho_{\text{W}}(\bm{x},\bm{p})= \frac{1}{\pi} \prod_{i} e^{-(x^{2}_{i}-p^{2}_{i})}.
\end{equation}
These $N_{\text{traj}}$ initial conditions are then propagated to a time $t_{\text{ss}}$ in which the vibrational steady state is reached, which we check by observing the time-evolution of the average kinetic energy of the vibrational modes. 

Once the steady state is reached, we propagate further and sample $N_{\text{sample}}$ equally separated points, for a total of $N_{\text{tot.}} = N_{\text{traj}} \times N_{\text{sample}}$ total points from which we calculate expectation values of observables. With this procedure, we can calculate the expectation values of adiabatic electronic observables as 
\begin{equation}\label{observable}
    \langle O \rangle^{\text{ss}}  = \frac{1}{N_{\text{tot.}}} \sum_{j = 1}^{N_{\text{tot.}}} \langle O \rangle_{\text{el}}(\bm{x}_{j}).
\end{equation}
Here, $\langle O \rangle_{\text{el}}(\bm{x})$ refers to the quantum expectation value of the electronic operator $O$ evaluated adiabatically at vibrational position $\bm{x}$ and in the corresponding electronic steady-state. For example, the adiabatic electrical current through lead $\alpha$ is \cite{PhysRevLett.68.2512, Preston2020}
\begin{align}
\langle I_\alpha \rangle_{\text{el}} (\bm{x}) = \: & \frac{1}{\pi}\int_{-\infty}^{\infty}d\epsilon \: \text{Re}\left(\text{Tr}\left\{ \tilde{G}_{(0)}^{<}\tilde{\Sigma}_{\alpha}^{A}+\tilde{G}_{(0)}^{R}\tilde{\Sigma}_{\alpha}^{<}\right\}\right). \label{eq: instantaneous adiabatic current}
\end{align}
One can use a similar expression to calculate expectation values of vibrational observables, the only difference being that $\langle O \rangle_{\text{el}}(\bm{x}_{j})$ are replaced by their classical vibrational counterparts. We are most interested in the steady-state excitation of the vibrational modes, $\langle N_{i} \rangle^{\text{ss}}$. Since we model our vibrational degrees of freedom as harmonic oscillators, we can approximate the excitation directly from the kinetic and potential energies as 
\begin{equation}\label{vib_exc_formula}
    \begin{aligned}
          \langle N_{i} \rangle^{\text{ss}} \approx \: & \frac{\langle E^{\text{ss}}_{\text{kin},i}\rangle + \langle E^{\text{ss}}_{\text{pot},i} \rangle}{\omega_{i}} =  \frac{1}{2} \left( \langle x _i^2 \rangle^{\text{ss}} + \langle p_i^2 \rangle^{\text{ss}} \right).
    \end{aligned}
\end{equation}

\subsection{Eigenmode Analysis} \label{subsec: Eigenmode Analysis}

Many previous investigations of vibrational instabilities via Langevin equations have not solved for the full stochastic dynamics directly, but rather reduced the problem to one solvable by an eigenmode analysis \cite{L2010, L2011, L2012}. In order to connect with these previous works and also to contrast with our direct solution, we will outline here how one can apply a similar approach to solve the Markovian Langevin equation we investigate. As we discuss in more detail in Sec.~\ref{subsec: Dynamics from an Eigenmode Analysis}, this is not the same eigenmode analysis as in previous work, as we start with a Markovian Langevin equation in which the electronic forces depend on the vibrational coordinates. 

We start by neglecting the stochastic force, such that the Langevin equation reduces to a deterministic differential equation
\begin{align}
    \ddot{x}_{i} \approx \: & -\omega_{i}^{2}x_{i} + F^{\text{ad}}_i(\bm{x}) - \sum_{j} \gamma_{ij}(\bm{x}) \dot{x}_{j}. \label{eq: deterministic langevin_equation}
\end{align}
At this point, we still cannot perform an eigenmode analysis of the problem, as the electronic forces are complicated functions of the vibrational coordinates. Therefore, the next step is to choose some point $\bm{x}_{0} = (x_{1,0},x_{2,0})$ around which we expand the forces. We seek a form of the adiabatic force linear in $\bm{x}$, so we approximate $\bm{F}^{\text{ad}}(\bm{x})$ as 
\begin{align}
    \begin{bmatrix}
         F_1^{\text{ad}} (\bm{x}) \\
         F_2^{\text{ad}} (\bm{x}) \\
     \end{bmatrix}
     \approx 
     \underbrace{\begin{bmatrix}
         F_1^{\text{ad}} (\bm{x}_{0}) \\
         F_2^{\text{ad}} (\bm{x}_{0}) \\
     \end{bmatrix}}_{= \bm{F}^{\text{ad},(0)}} + \underbrace{\begin{bmatrix}
      \partial_{1} F_1^{\text{ad}} |_{\bm{x}_{0}} & \partial_{1} F_1^{\text{ad}}|_{\bm{x}_{0}}  \\
      \partial_{2} F_2^{\text{ad}}|_{\bm{x}_{0}} & \partial_{2} F_2^{\text{ad}}|_{\bm{x}_{0}}
     \end{bmatrix}}_{= \bm{F}^{\text{ad},(1)}}
     \begin{bmatrix}
     x_{1} \\ x_{2}
      \end{bmatrix}, \label{force_vec_linearized}
\end{align}
Given that the constant force $\bm{F}^{\text{ad},(0)}$ does not affect the long-time dynamics, we will neglect it from here on. 

Using the same expansion point, we could also linearize the electronic friction. However, since an eigenmode analysis requires an equation of motion linear in the vibrational velocities, we will actually take the zeroth-order contribution instead:
\begin{align}
    \boldsymbol{\gamma}(\bm{x})  \approx \: & \boldsymbol{\gamma}(\bm{x}_{0}) = \boldsymbol{\gamma}^{(0)},
\end{align}
such that the deterministic approximation to the Langevin equation linearized at the point $\bm{x}_{0}$ is 
\begin{align}
    \Bigg[ \begin{array}{c}
    \ddot{x}_1 \\ \ddot{x}_2
    \end{array} \Bigg]
    = \: & \left[ \begin{array}{cc}
    -\omega_{1}^{2} + F_{11}^{\text{ad},(1)} & F_{12}^{\text{ad},(1)} \\
    F_{21}^{\text{ad},(1)} & -\omega_{2}^{2} +F_{22}^{\text{ad},(1)}
    \end{array} \right]
    \Bigg[ \begin{array}{c}
    x_1 \\ x_2
    \end{array} \Bigg] \nonumber \\
    & - \Bigg[ \begin{array}{cc}
    \gamma^{(0)}_{11} & \gamma^{(0)}_{12} \\
    \gamma^{(0)}_{21} & \gamma^{(0)}_{22}
    \end{array} \Bigg]
    \Bigg[ \begin{array}{c}
    \dot{x}_{1} \\ \dot{x}_{2}
    \end{array} \Bigg], \label{langevin_equation_without_random_force_x0_y0}
\end{align}
We now proceed to solve Eq.\eqref{langevin_equation_without_random_force_x0_y0} with the ansatz 
\begin{align}
    \bm{x}(t) = \: & \text{Re}\left\{\bm{u} e^{i\Omega(\bm{x}_{0})t}\right\} \label{ansatz_Omega}
\end{align}
where $\Omega(\bm{x}_{0})$ is the frequency of an eigenmode depending on the choice of the expansion point, $\bm{x}_{0}$, and $\bm{u}$ is the corresponding eigenvector, normalized such that \\ $\sum\limits_{i = 1}^{2}u_{i}=1$. Inserting Eq.\eqref{ansatz_Omega} in Eq.\eqref{langevin_equation_without_random_force_x0_y0}, we obtain 
\begin{align} \label{eigenvector_equation}
    \mathcal{K}(\Omega) \bm{x} = \: & -\Omega^{2}\bm{x}
\end{align}
where the dynamical matrix is 
\begin{align} 
    \mathcal{K}(\Omega) = \: & \begin{bmatrix}
         -\omega_{1}^2 - i\Omega \gamma_{11} + F^{\text{ad},(1)}_{11} & -i \Omega \gamma_{12} + F^{\text{ad},(1)}_{12}
         \\
         -i\Omega\gamma_{21} + F^{\text{ad},(1)}_{21} & -\omega_{2}^2 - i\Omega\gamma_{22}+F^{\text{ad},(1)}_{22}
    \end{bmatrix}.
     \label{system_linear_equation}
\end{align}
We rewrite the two vibrational frequencies by renaming $\omega_{1} = \omega$ and defining $\omega_{2}$ in terms of some detuning, $\Delta = \omega_{1} - \omega_{2}$, which in our work is always much smaller than the frequencies themselves: $\Delta \ll \omega$. Furthermore, in the limit of small damping due to electronic friction, the eigenmodes must be close to the undamped frequencies, $\Omega \approx \omega$, such that $\mathcal{K}(\Omega) \approx \mathcal{K}$ and 
\begin{align} 
    \mathcal{K} \approx \: & \begin{bmatrix}
         -\omega^2 - i\omega \gamma_{11} + F^{\text{ad},(1)}_{11} & -i \omega \gamma_{12} + F^{\text{ad},(1)}_{12}
         \\
         -i\omega\gamma_{21} + F^{\text{ad},(1)}_{21} & -(\omega - \Delta)^2 - i\omega\gamma_{22}+F^{\text{ad},(1)}_{22}
    \end{bmatrix}.
     \label{eq: system_linear_equation limited}
\end{align}
As a result, in this limit we can combine Eq.\eqref{eq: system_linear_equation limited} with Eq.\eqref{eigenvector_equation}  to calculate two eigenmodes for the deterministic problem, and their corresponding eigenfrequencies, $\Omega_{\nu}(\bm{x}_{0})$, which will depend on the choice of expansion point, $\bm{x}_{0}$. Note that formally we would obtain four possible solutions for $\Omega_{\nu}$ but that we only retain the two solutions satisfying $\text{Re}( \Omega_{\nu})>0$.

As a measure of instability in the eigenmode picture, we will use the inverse Q-factor,
\begin{align}
1/Q_{\nu}(\bm{x}_{0}) = \: & -2 \frac{\text{Im} \left(\Omega_{\nu}(\bm{x}_{0})\right)}{\text{Re} \left(\Omega_{\nu}(\bm{x}_{0})\right)}, \label{eq: inverse q factor}
\end{align}
which describes the rate at which energy is lost from the eigenmode. If the inverse Q-factor of an eigenmode is negative $1/Q_{\nu} < 0$, this implies that $\bm{x}_{0}$ is a point in space at which energy is being pumped into eigenmode $\nu$. This approach to understanding instabilities in molecular junctions is potentially highly useful and has been previously explored for two-level, two-mode systems. Particularly interesting are reports of so-called \textit{runaway modes}, where the inverse Q-factor of one eigenmode is positive (stable) while the inverse Q-factor of the other eigenmode is negative (unstable), as these would be experimentally accessible signatures for instabilities arising from nonconservative current-induced forces \cite{L2010,L2011,L2012}.

\section{Results}\label{sec: Results}

\noindent In this section, we will investigate a model system with two electronic states and two vibrational modes for which vibrational instabilities have previously been reported~\cite{L2011}. In the first part of this section, we perform full Langevin simulations of the vibrational dynamics, analyzing the influence of different electronic forces and the occurrence of vibrational instabilities. In the second part, we perform an eigenmode analysis motivated by the work in Refs.~\cite{L2010, L2011} and investigate the reliability of such approaches in predicting vibrational instabilities, especially for approaches in which the electronic forces depend on the vibrational coordinates. 

\subsection{Nonequilibrium Transport Observables in the Steady-State} \label{subsec: Directly Solving the Langevin Equation}

We first discuss steady-state expectation values of observables for the two-level, two-mode model in Eq.\eqref{eq: molecular Hamiltonian}, calculated by numerically solving the full Langevin equation in Eq.\eqref{langevin_equation} with the algorithm outlined in Sec.~\ref{subsec: Langevin simulations and Observables}. Specifically, we are interested in investigating regimes in which there have previously been reports of instabilities arising from nonconservative current-induced forces. To this end, we focus on the parameters outlined in Tbl.~\ref{tab: paras_table}. Previous mixed quantum-classical investigations of these parameters have predicted strong instabilities at bias voltages much lower than that predicted solely by current-induced heating \cite{L2010,L2011,L2012}. Since these analyses have largely relied on an eigenmode picture similar to that outlined in Sec.~\ref{subsec: Eigenmode Analysis}, a natural question is whether these instabilities are also present in the full treatment.

\begin{table}[h]
\centering
\setlength{\tabcolsep}{0pt}
\renewcommand{\arraystretch}{1.5} 
\begin{tabularx}{\columnwidth}{@{} >{\centering\arraybackslash}p{0.5\columnwidth} >{\centering\arraybackslash}p{0.5\columnwidth}}
\rowcolor{gray!20}
\textbf{Vibrational Parameters} & {} \\ 
$\omega_{0}$ & $20\text{ meV}$\\
$\Delta$ & $0-5\text{ meV}$\\
$\omega_{i}$ & $\omega_{0} \pm \Delta/2$\\
\rowcolor{gray!20}
\textbf{Electronic Parameters} & {} \\ 
$\varepsilon_{1}$ & $0\text{ meV}$\\
$\varepsilon_{2}$ & $0\text{ meV}$\\
$t$ & $-0.2\text{ meV}$\\
$k_{B}T$ & $25.8\text{ meV}$\\
$\Gamma_{\alpha}$ & $1\text{ eV}$\\
\rowcolor{gray!20}
\multicolumn{2}{p{1\columnwidth}}{\textbf{Electronic-Vibrational Interaction Parameters}} \\
$\lambda_{1}$ & $4.3\text{ meV}$ \\ 
$\lambda_{2}$ & $4.18\text{ meV}$ \\
\end{tabularx}
\caption{Vibronic parameters for the two-level, two-mode model introduced in Eq.\eqref{eq: molecular Hamiltonian}.}
\label{tab: paras_table}
\end{table}

\begin{figure} 
     \centering
     \includegraphics[width=\columnwidth, trim=10 7 10 10, clip]{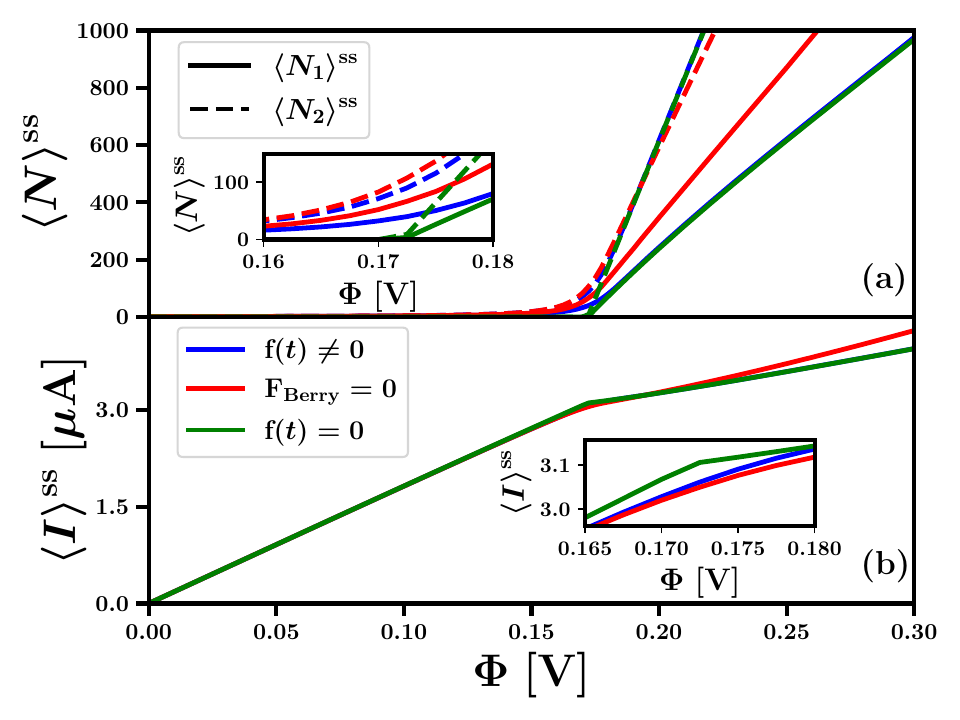}
          \phantomsubfloat{\label{fig: full transport observables zero delta vibrational excitation}}
     \phantomsubfloat{\label{fig: full transport observables zero delta current}}
        \vspace{-0.9cm}
     \caption{Steady-state expectation values of transport observables as a function of bias voltage for degenerate vibrational frequencies: $\Delta = 0$. In (a), the excitation of both vibrational modes is shown, $\langle N_{i} \rangle^{\text{ss}}$, while (b) contains the electrical current through the left lead, $\langle I_{L} \rangle^{\text{ss}}$. Blue lines correspond to the full solution of the Langevin equation, while red lines correspond to turning off the Berry force $F_{\text{Berry}}$ by setting $\gamma=\gamma^{\text{S}}$ (see Eq.\eqref{eq:friction_sym_antisym}). The green lines correspond to the deterministic solution when the stochastic force is removed: $\mathbf{f}(t) = \boldsymbol{0}$. Other parameters are listed in Tbl.~\ref{tab: paras_table}. Note that in (b) the green line and the blue line lie almost on top of one another for $\Phi>0.18~\text{V}$.}
     \label{fig: full transport observables zero delta}
 \end{figure}

We start with the case of degenerate modes, $\Delta = 0$, for which the steady-state transport observables as a function of bias voltage are shown in Fig.~\figref{fig: full transport observables zero delta}. First, we will analyze the vibrational excitation, shown in Fig.~\figref{fig: full transport observables zero delta vibrational excitation}. We see that there \textit{is} indeed a vibrational instability for degenerate vibrational frequencies. The excitation of both vibrational modes exceeds $100$ at very low bias voltages, $0.15\text{ V} \leq \Phi \leq 0.2 \text{ V}$, and continues to grow as a function of bias voltage well beyond the limits of what is considered stable for a molecular nanojunction. Note that this is not evidence of a runaway mode, as we have plotted the vibrational excitations of the raw vibrational modes, not the eigenmodes of the problem. 

This instability has a different origin than simple current-induced or Joule heating. First, one sees that it occurs at a much lower bias voltage than that which is typical for current-induced heating, and, second, that it is not influenced by the inclusion of the stochastic forces. This is shown in the green lines of Fig.~\figref{fig: full transport observables zero delta vibrational excitation}, which are the transport observables obtained by solving the deterministic equation resulting from setting $\mathbf{f}(t) = \mathbf{0}$ in Eq.\eqref{langevin_equation}, and lie almost directly on top of the observables obtained from the full calculation. As discussed in Sec.~\ref{subsec: Electronic Friction and Langevin Dynamics}, in the NEGF-LD picture, Joule heating manifests itself in nonequilibrium charge transport as a breaking of the fluctuation-dissipation theorem at finite bias voltages and transfer of energy from electrical current to the vibrations \cite{Rudge2023,Rudge2024}. For these parameters, however, we observe that removing the stochastic force does not decrease the vibrational excitation, indicating that the instability \textit{does not} arise from current-induced heating. 

As discussed in Refs.~\cite{L2011,L2012}, the mechanism for this instability is instead the nonconservative current-induced forces that arise even at low bias voltages. Indeed, one observes from the green lines of Fig.~\figref{fig: full transport observables zero delta vibrational excitation} that for voltages below $\Phi \approx 0.17 \text{ V}$, the vibrational excitation in the steady state is zero. In this voltage range, the instability caused by nonconservative current-induced forces is not present and the vibrational modes continuously lose energy due to the damping caused by the electronic friction, which the stochastic force cannot compensate since it has been artificially turned off. When the instability due to the nonconservative electronic forces turns on, then, we observe a kink in the steady-state expectation values of observables. Finally, we observe that the instability arises not just from the nonconservative current-induced forces, but specifically from the nonconservative part of $F_{\text{ad}}(\bm{x})$. This is shown by the red lines of Fig.~\figref{fig: full transport observables zero delta vibrational excitation}, which we obtained by removing the antisymmetric part of the friction tensor and leaving only the symmetric, dissipative part, $\gamma^{S}$. 

A key concern surrounding mechanical instabilities in molecular junctions is that they can have a significant impact on the conductance, which we explore in Fig.~\figref{fig: full transport observables zero delta current} via the steady-state current. Despite the high vibrational excitation, the corresponding steady-state current remains stable. Since the molecule-lead coupling is large, $\Gamma = 1\text{ eV}$, the current does not display step-like behavior when a new electronic transport channel opens, but rather increases steadily as a function of bias voltage. Additionally, since the vibrational modes are treated classically and, thus, can essentially continuously exchange energy with transporting electrons, the current also does not display any steps due to vibrational quantization \cite{Rudge2024}. We note that this behavior is atypical for a nanojunction experiencing instability; one would instead expect that such large vibrational excitations would cause the junction to break, which is not incorporated in this harmonic model. More realistic models would have anharmonic potentials, allowing for bond rupture, and a position-dependent molecule-lead coupling, such that the conductance decreases for high vibrational excitation \cite{Ke2021,Ke2023,Erpenbeck2023,Erpenbeck2020,Erpenbeck2019}. This remains the focus of future work. 

\begin{figure} 
     \centering
     \includegraphics[width=\columnwidth, trim=10 0 10 10, clip]{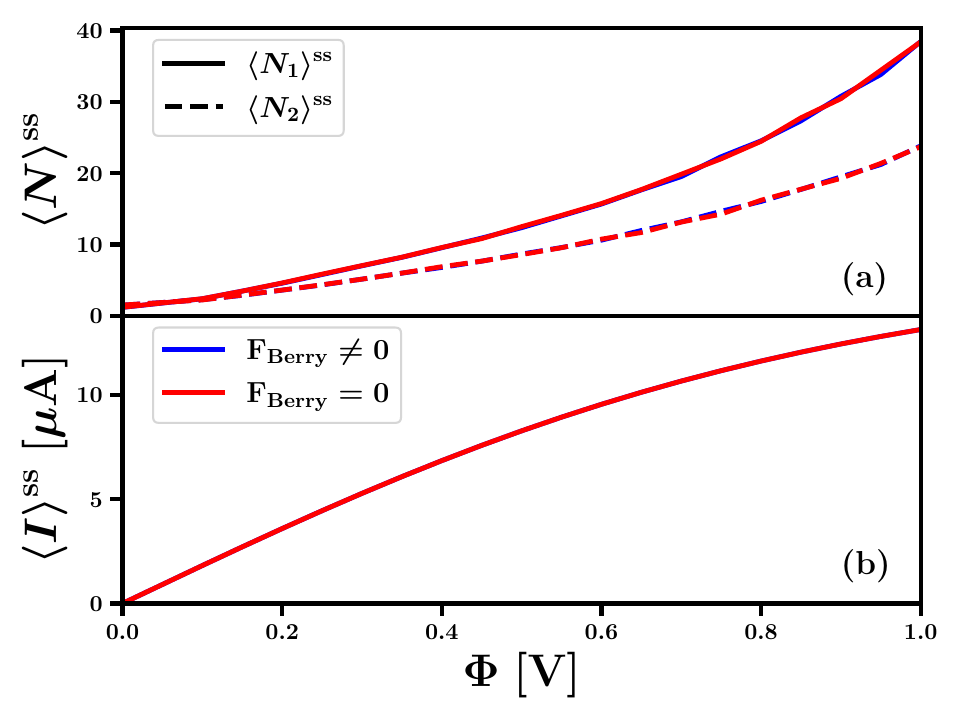}
     \phantomsubfloat{\label{fig: full transport observables nonzero delta vibrational excitation}}
     \phantomsubfloat{\label{fig: full transport observables nonzero delta current}}
     \vspace{-1cm}
     \caption{Similar analysis as in Fig.~\figref{fig: full transport observables zero delta}, except with detuned vibrational frequencies: $\Delta = 5\text{ meV}$. Other parameters are the same as in Fig.~\figref{fig: full transport observables zero delta}. Note that a larger voltage range has been used than in Fig.~\figref{fig: full transport observables zero delta}, as the dynamics remains stable.}
     \label{fig: full transport observables nonzero delta}
 \end{figure}

Next, we consider the case of nondegenerate modes, $\Delta = 5\text{ meV}$, shown in the blue lines of Fig.~\figref{fig: full transport observables nonzero delta}. In contrast to the case of degenerate frequencies, the steady-state dynamics are now much more stable. The vibrational excitation of both modes remains relatively small, even in the high-bias regime: $\langle N_{i} \rangle^{ss} \leq 40$. Although not shown here, the steady-state vibrational excitation reduces to zero without the inclusion of the stochastic force, $\mathbf{f}(t) = \mathbf{0}$, indicating that current-induced heating is the dominant effect. Evidently, the mechanism of vibrational instability arising from nonconservative current-induced forces disappears when the vibrational modes are detuned. 

The stability of the vibrational dynamics also clearly affects the current, which is seen in Fig.~\figref{fig: full transport observables nonzero delta current}. Here, $\langle I \rangle^{\text{ss}}$ is consistently larger than its counterpart for the case of degenerate modes in Fig.~\figref{fig: full transport observables zero delta current}. The instantaneous adiabatic current, $\langle I_\alpha \rangle_{\text{el}}(\bm{x})$, depends on the electronic eigenenergies, which in turn are functions of the vibrational coordinates: $E_{\pm}(\bm{x})$. At vibrational positions where the electronic energies sit inside (outside) the bias window, $\langle I_\alpha \rangle_{\text{el}}(\bm{x})$ is large (small). Since a large vibrational excitation corresponds to a large variance in the vibrational coordinates,
\begin{align}
(\Delta x_{i})^{2} = \: & \langle x_{i}^{2}\rangle^{\text{ss}} - (\langle x_{i}\rangle^{\text{ss}})^{2} \gg 1,
\end{align}
in regimes of high vibrational excitation the electronic eigenenergies are, on average, located outside the bias window and the corresponding average current is smaller. 

Finally, in this subsection, we note that our result for the detuned case contradicts previous investigations of this model, which found instabilities in the form of runaway modes even for $\Delta = 5\text{ meV}$ \cite{L2010,L2011,L2012}. The key difference to our results is that these investigations relied on a similar eigenmode analysis as that outlined in Sec.~\ref{subsec: Eigenmode Analysis}, albeit with a weak-coupling as opposed to Markovian approximation for the vibrational dynamics. The proposed mechanism relied on the Berry force, arguing that at finite detunings it was critical in pulling the vibrational dynamics into elliptical rather than linear trajectories, allowing the nonconservative current-induced adiabatic force to produce a runaway vibrational eigenmode, similarly to the case of $\Delta = 0$. However, not only do we not observe a vibrational instability at these parameters, but we also only observe a minimal influence of the Berry force on the detuned vibrational dynamics. This is shown in the red lines of Fig.~\figref{fig: full transport observables nonzero delta}, where the steady-state expectation values of the vibrational excitation and current are almost identical between the dynamics with and without the Berry force. 

\subsection{Eigenmode Analysis} \label{subsec: Dynamics from an Eigenmode Analysis}

In this subsection, our goal is to analyze the vibrational dynamics from the simplified eigenmode perspective, comparing and contrasting this with the dynamics obtained from the full solution of the Langevin equation. To this end, we will analyze the same parameters and detunings as in Sec.~\ref{subsec: Directly Solving the Langevin Equation} using the approach outlined in Sec.~\ref{subsec: Eigenmode Analysis}.

\begin{figure*} 
\begin{center}
     \centering     
     \includegraphics[width=1.0\textwidth, trim=67 0 77 40, clip]{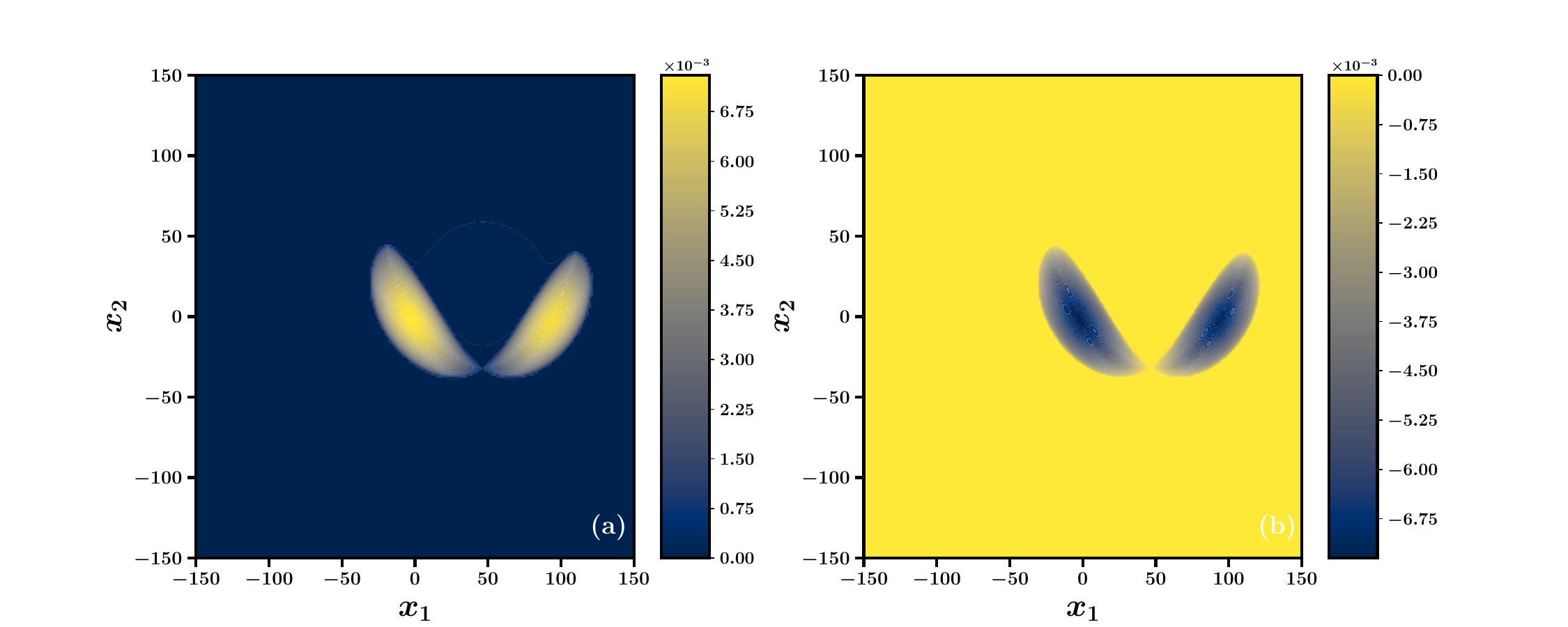}
     \phantomsubfloat{\label{fig: iqf_mode_1_color_plot_delta_0.0ev_voltage_0.2V}}
     \phantomsubfloat{\label{fig: iqf_mode_2_color_plot_delta_0.0ev_voltage_0.2V}}
     \vspace{-.80cm}
     \caption{Inverse Q-factor, $1/Q_{\nu}$, for both eigenmodes as a function of the expansion point, $\bm{x}_{0}$. The detuning is $\Delta = 0$, the voltage is $\Phi=0.2$~\text{V}, and all other parameters are the same as in previous plots.}
     \label{fig: iqf_color_plot_delta_0.0ev_02v}
     \end{center}
 \end{figure*}   

Again, we start with the degenerate-mode case, shown in Fig.~\figref{fig: iqf_color_plot_delta_0.0ev_02v}, where we have plotted the inverse Q-factor, $1/Q_{\nu}$, for both eigenmodes as a function of the expansion point, $\bm{x}_{0}$. The inverse Q-factor measures the rate at which energy is lost from eigenmode $\nu$. As shown in Sec.~\ref{subsec: Eigenmode Analysis} the $\nu$th eigenmode is a solution to the deterministic equation of motion obtained by neglecting the stochastic force and linearizing the electronic forces at coordinate $\bm{x}_{0}$. Consequently, if the $\nu$th inverse Q-factor is negative, it implies that the $\nu$th eigenmode of the deterministic dynamical problem in the immediate vicinity of $\bm{x}_{0}$ is unstable, as this mode will gain energy per cycle rather than lose it.

This is what we observe for eigenmode $1$ in Fig.~\figref{fig: iqf_mode_1_color_plot_delta_0.0ev_voltage_0.2V}, where the inverse Q-factor is either close to zero or negative for the entire vibrational coordinate space. In regions where the inverse Q-factors are zero the current-induced forces have a negligible impact on the dynamics and the problem becomes undamped: $\text{Im}\{\Omega_{\nu}\} \rightarrow 0$. Consequently, at this voltage the deterministic approximation to the full dynamics accurately predicts a vibrational instability as this eigenmode is either undamped or unstable for the entire coordinate space. Although not shown here, the inverse Q-factors for all voltages greater than $\Phi \geq 0.2~\text{ V}$ also display this behavior. 

\begin{figure} 
     \centering
     \includegraphics[width=\columnwidth, trim=10 0 10 10, clip]{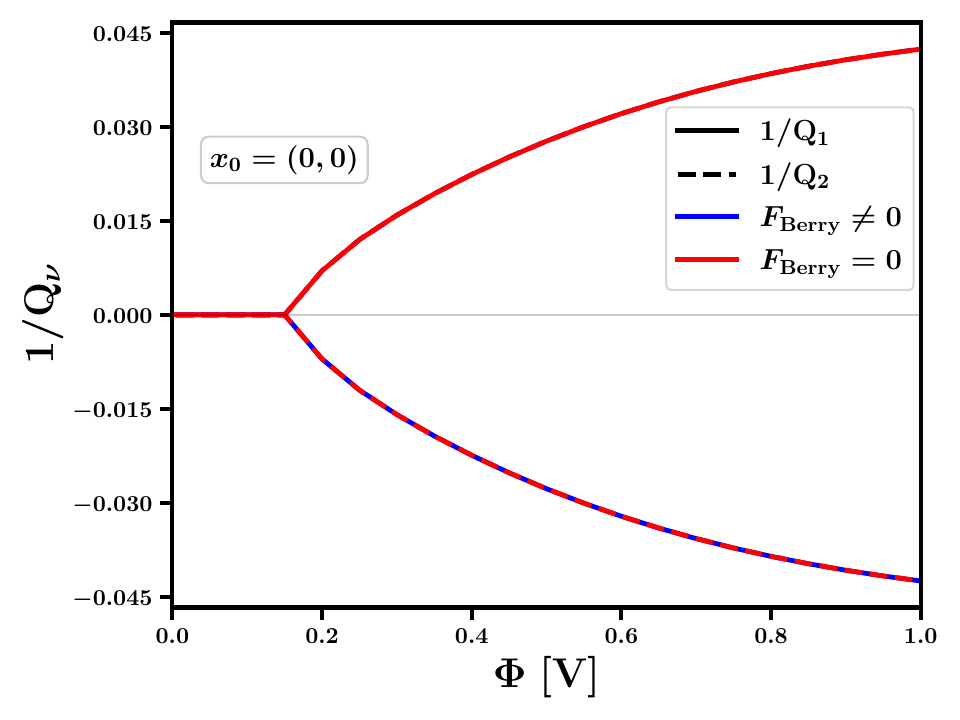}
     \vspace{-0.75cm}
     \caption{Inverse Q-factor, $1/Q_{\nu}$, calculated at $\bm{x}_{0}=(0,0)$, for different bias voltages. 
     The gray line is at $1/Q_{\nu}=0$. Blue lines include the Berry force, while red lines correspond to removing the Berry force. 
     The detuning is $\Delta = 0$. All other parameters are the same as in previous plots.}
     \label{fig:inverse_Qfac_different_voltages_w0_0.02ev_delta_0}
 \end{figure}

Next, consider what would happen to our analysis if we were to pick a single point and approximate the forces for the entire vibrational coordinate space as those expanded around. In Fig.~\figref{fig:inverse_Qfac_different_voltages_w0_0.02ev_delta_0}, we show the inverse $Q$-factor for different bias voltages at the expansion point $\bm{x}_{0}=(0,0)$. First, we note that the Berry force has no visible impact on the onset of the runaway behavior for all voltages, indicating again that the dominant effect on the instability arises from the nonconservative part of the adiabatic contribution to the mean force. 

Second, the inverse $Q$-factor of eigenmode $2$ becomes negative for voltages $\Phi>0.17~\text{V}$. Based on this analysis, this mode should display runaway behavior for bias voltages $\Phi>0.17~\text{V}$. This runaway mode evidently translates to an instability in the full vibrational dynamics, which we see when comparing with the results shown in Fig.~\figref{fig: full transport observables zero delta vibrational excitation}. There, we observed an immense excitation of both of the \textit{raw} vibrational modes at this bias voltage.

Each choice of $\bm{x}_{0}$ represents a different eigenmode, formed by a different linear combination of the $x_{\nu}$. Therefore, when one runs the full deterministic calculations, the long-time dynamics is some mixture of the dynamics approximated at each point in the vibrational coordinate space. Since all choices of $\bm{x}_{0}$ 
produce either undamped or runaway modes, the instability observed in the individual $1/Q_{\nu}$ extends to the raw vibrational modes and we conclude that the eigenmode analysis is reliable for degenerate modes, $\Delta = 0$. However, in regimes where $1/Q_{\nu}$ is positive for some regions of the vibrational coordinate space and negative for others,  $\bm{x}_{0}$ and negative for others, such an analysis is difficult to apply. 

\begin{center}
    \begin{figure*} 
         \centering     
         \includegraphics[width=\textwidth, trim=67 0 77 40, clip]{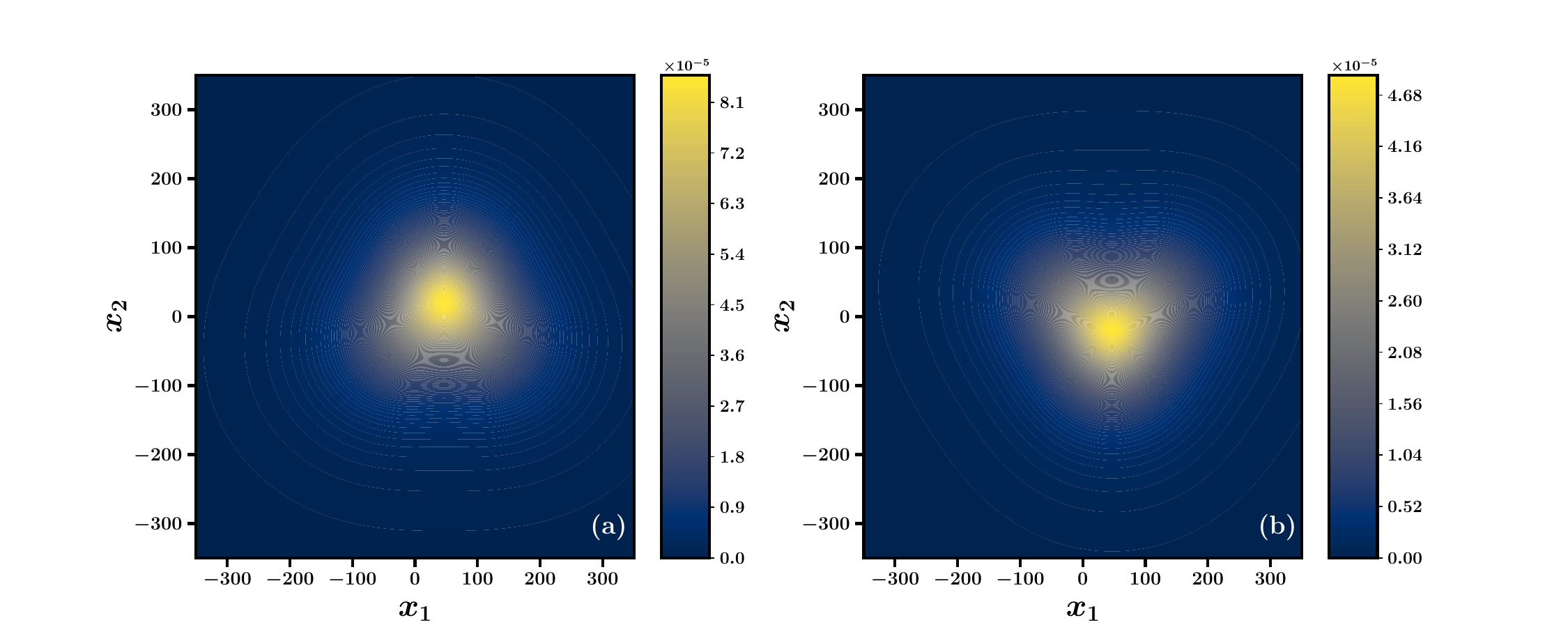}
         \includegraphics[width=\textwidth, trim=67 0 77 40, clip]{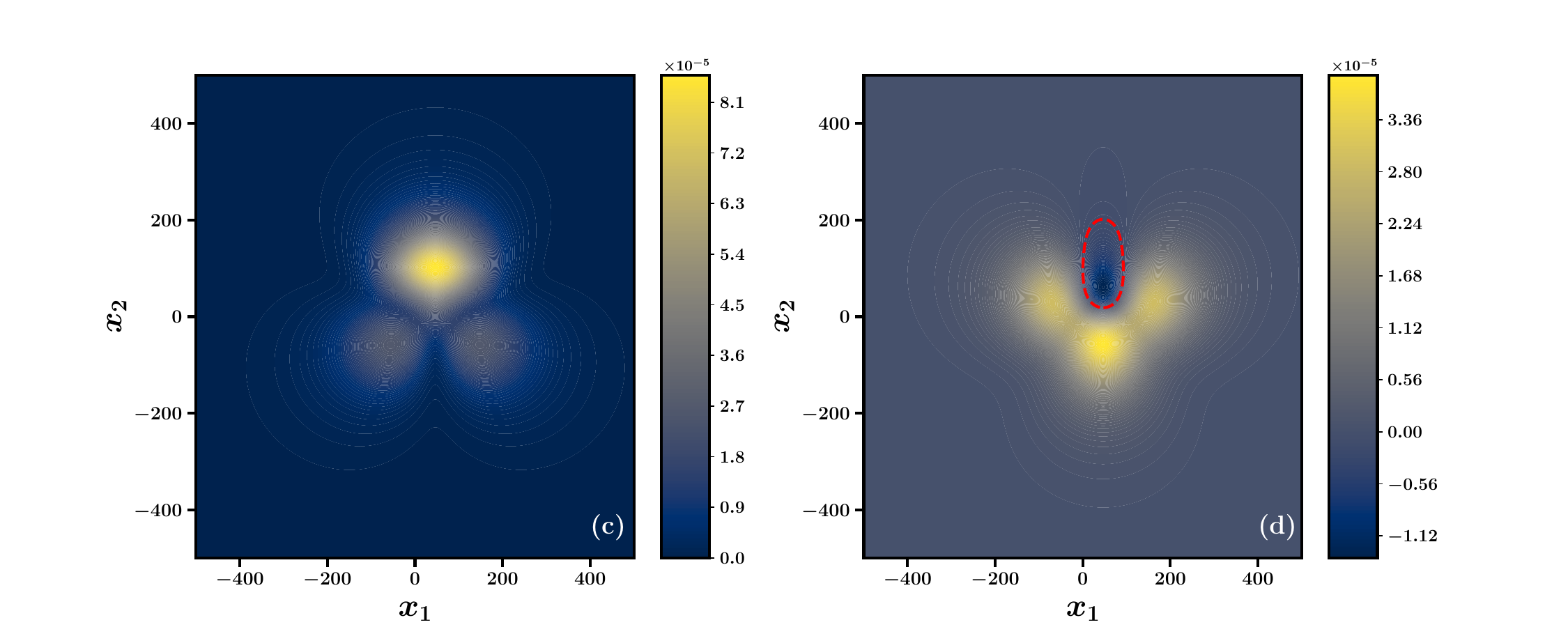}
         \phantomsubfloat{\label{fig: iqf_mode_1_color_plot_delta_0.005ev_voltage_0.2v}}
         \phantomsubfloat{\label{fig: iqf_mode_2_color_plot_delta_0.005ev_voltage_0.2v}}     
         \phantomsubfloat{\label{fig: iqf_mode_1_color_plot_delta_0.005ev_voltage_1v}}
         \phantomsubfloat{\label{fig: iqf_mode_2_color_plot_delta_0.005ev_voltage_1v}}
         \vspace{-0.75cm}
         \caption{Inverse Q-factor, $1/Q_{\nu}$, for both eigenmodes as a function of the expansion point, $\bm{x}_{0}$, and at two different voltages. The detuning is $\Delta = 5\text{ meV}$ and all other parameters are the same as in previous plots. The upper row, (a) and (b), are calculated at $\Phi = 0.2\text{ V}$, while the bottom row, (c) and (d) are calculated at $\Phi = 1\text{ V}$. 
         The regions where $1/Q_{\nu}$ becomes negative have been highlighted by a red dashed line.
         Note that in (c) and (d) a larger grid has been used compared to (a) and (b).}
         \label{fig: iqf_color_plot_delta_0.005ev}
    \end{figure*}
\end{center}

This is exactly what we observe for the case of finite detuning, $\Delta = 5\text{ meV}$, which is shown in Fig.~(\ref{fig: iqf_color_plot_delta_0.005ev}). In the upper row, at a voltage of $\Phi = 0.2\text{ V}$, the inverse Q-factors of both eigenmodes are positive for all $\bm{x}_{0}$, suggesting that the eigenmodes and thus the raw vibrational modes are damped. This is in agreement with the vibrational excitation of the individual vibrational modes calculated from the full dynamics in Fig.~\figref{fig: full transport observables nonzero delta}, where we observed $\langle N_{i} \rangle^{\text{ss}} < 5$ at this bias voltage. However, the situation changes when we increase the bias voltage to $\Phi = 1\text{ V}$, which is shown in Figs.\figref{fig:  iqf_mode_1_color_plot_delta_0.005ev_voltage_1v}-\figref{fig:  iqf_mode_2_color_plot_delta_0.005ev_voltage_1v}. Here, $1/Q_{{1}}$ remains positive for the entire grid, but $1/Q_{{2}}$ is negative for a certain region in the vicinity of $\bm{x}=\left(50,55\right)$, which we have highlighted with a dashed red border.

\begin{figure} 
     \centering
     \includegraphics[width=\columnwidth, trim=10 0 10 10, clip]{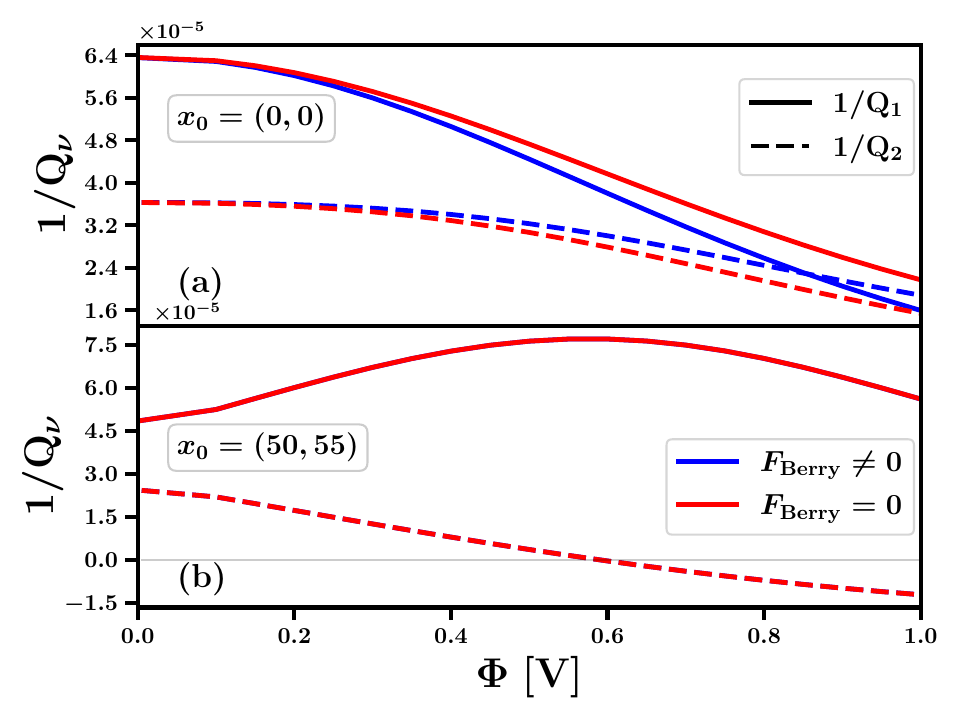}
          \phantomsubfloat{\label{fig:inverse_Qfac_different_voltages_w0_0.02ev_delta_0.005ev_a}} \phantomsubfloat{\label{fig:inverse_Qfac_different_voltages_w0_0.02ev_delta_0.005ev_b}}
     \vspace{-1.15cm}
     \caption{Inverse Q-factor, $1/Q_{\nu}$, at different expansion points, $\bm{x}_{0}=(0,0)$ (a) and $\bm{x}_{0}=(50,55)$ (b), for different bias voltages. 
     The gray line in (b) is at $1/Q_{\nu}=0$. Blue lines include the Berry force, while red lines correspond to the removing the Berry force. 
     The detuning is $\Delta = 5~\text{meV}$. All other parameters are the same as in previous plots.}
     \label{fig:inverse_Qfac_different_voltages_w0_0.02ev_delta_0.005ev}
 \end{figure}

Here, eigenmode $2$ is damped by current-induced electronic forces when the vibrational dynamics is outside the zone outlined in red, but negatively damped and driven to instability when the vibrational dynamics is within the red region. Considering that in a dynamical calculation, there will be times for which the vibrational dynamics is in this region and also times for which the vibrational dynamics is outside, it is hard to predict \textit{a priori} from the eigenmode analysis alone whether this is a runaway mode and whether it translates into an instability in the raw vibrational modes. 

In Fig.~\figref{fig:inverse_Qfac_different_voltages_w0_0.02ev_delta_0.005ev}, we show the inverse $Q$-factors for different bias voltages, where the electronic forces have been approximated from two different expansion points. The first expansion point, $\bm{x}_{0} = (0,0)$, lies outside the region of negativity in Fig.~\figref{fig: iqf_mode_2_color_plot_delta_0.005ev_voltage_1v}, predicting that eigenmode $2$ is damped for all bias voltages, and, thus, stable under this approximation. In contrast, the second expansion point, $\bm{x}_{0} = (50,55)$, lies inside the region with negative inverse $Q$-factor in Fig.~\figref{fig: iqf_mode_2_color_plot_delta_0.005ev_voltage_1v}. One observes that, for this expansion point, there is a critical voltage, $\Phi_{\text{crit.}} \approx 0.6 \text{ V}$ at which $1/Q_2$ becomes negative and the dynamics would exhibit a runaway mode. Evidently, then, a simple eigenmode analysis yields conflicting predictions for the vibrational stability depending on the choice of the expansion point. Note that also in Fig.~\figref{fig:inverse_Qfac_different_voltages_w0_0.02ev_delta_0.005ev} the Berry force has no significant impact on the stability of the system. 

We now compare this to the results of the full dynamics from Fig.~\figref{fig: full transport observables nonzero delta}, where we observed only a moderate vibrational excitation for finite detuning as a function of bias voltage. Unlike the case of degenerate modes, it is clear that here we cannot use the eigenmode analysis as conclusive evidence of vibrational instabilities. The vibrational trajectories might not spend enough time in the region of vibrational coordinates where the inverse Q-factor is negative; they might be strongly damped in other parts of the grid; or they might not even reach this part of the grid at all. This example shows that this simple eigenmode analysisi, in which the full nonlinear, coordinate-dependent dynamics is approximated at a single vibrational point, are insufficient in describing the complexity of the full nonequilibrium problem. 

We will end this subsection with a discussion of the connections between our results and previous investigations of this model \cite{L2012}, which also relied on an eigenmode analysis to predict the appearance of runaway modes. First, we note that most investigations of charge transport through two-level, two-mode models relying on current-induced forces have used a slightly different version of the Langevin equation, derived under different approximations \cite{L2010,L2011,L2012,Chen2018,Chen2019}. Specifically, if one assumes weak nonadiabaticity by enforcing the limit of small electronic-vibrational coupling rather than the adiabatic approximation, one obtains a generalized Langevin equation,
\begin{align}
    m_{i} \ddot{x}_{i} = \: & - \omega_{i} x_{i} - \sum_{j}\int^{t}_{0} d\tau \: {\Pi}_{ij}(t - \tau) x_{j}(\tau) + f_{i}(t),
\end{align}
where the memory kernel, $\bm{\Pi}(t - \tau)$, does not depend on the vibrational coordinates. Under a Fourier transform, the memory kernel can be split into the friction, the Berry force, and a nonconservative force proportional to the vibrational coordinate. Crucially, these forces are often approximated at the Fermi level only, reducing the problem to the same form as that given in Eq.\eqref{langevin_equation_without_random_force_x0_y0}, which can be treated via an eigenmode analysis. The key here is that, although the approximations and Langevin equation are slightly different than in our approach, previous investigations have also relied on a simplification of the full current-induced electronic forces to make predictions about the stability of the junction, in this case by taking only a single frequency. As we have shown above, it appears that such analyses can predict different dynamics than what is obtained from the full Langevin simulation.  

\subsection{Influence of Vibrational Detuning on Instabilities}

In the previous two subsections, we have shown that the instability induced by nonconservative current-induced forces at zero detuning disappears for a detuning of at least $\Delta = 5\text{ meV}$. In this subsection, we investigate exactly how sensitive the mechanism for these instabilities is to the detuning. To this end, we simulate the full Langevin dynamics for the same parameters as in Section~\ref{subsec: Directly Solving the Langevin Equation}, but now for a variety of $\Delta$. 

\begin{figure} 
     \centering
     \includegraphics[width=\columnwidth, trim=10 0 10 10, clip]{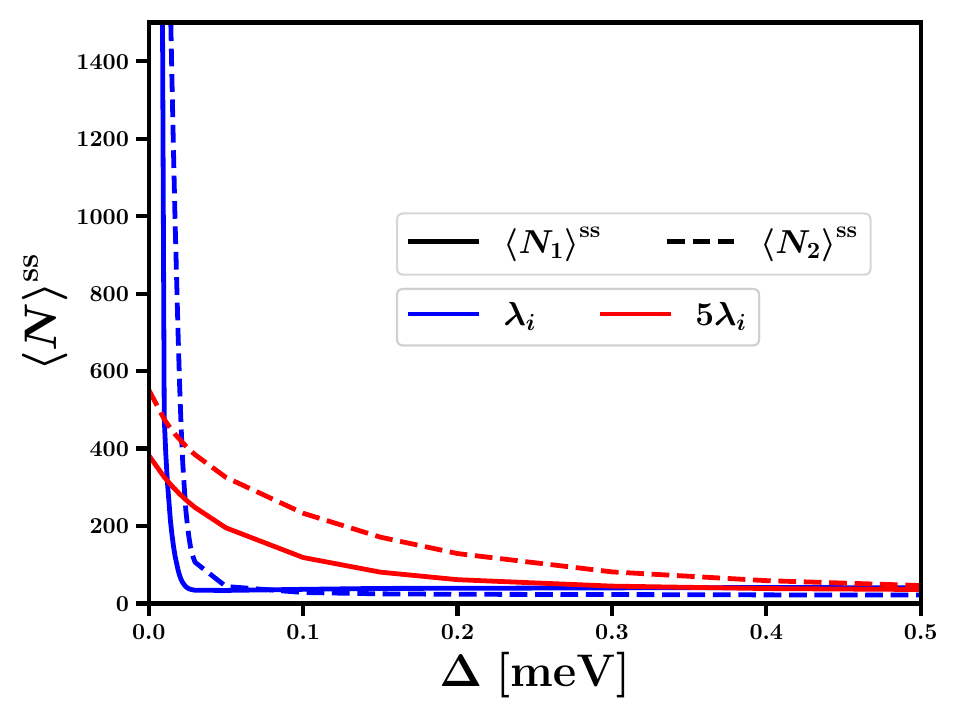}
     \vspace{-0.75cm}
     \caption{Vibrational excitation of both modes for different detunings. For the blue lines the same electronic-vibrational couplings $\lambda_i$ as listed in Table~\ref{tab: paras_table} have been used, while for red lines both couplings have been multiplied by a factor of 5. The voltage is $\Phi=1\text{V}$. Note that the largest detuning shown here is still a factor of 10 smaller than the detuning used in Section~\ref{subsec: Directly Solving the Langevin Equation}.}
     \label{fig: phonon_number_different_detunings}
 \end{figure}

In Fig.~\figref{fig: phonon_number_different_detunings}, we show the vibrational excitation of both modes calculated for different detunings and different electronic-vibrational couplings at a voltage of $\Phi=1~\text{V}$.
For the electronic-vibrational coupling used in Section~\ref{subsec: Directly Solving the Langevin Equation} (blue lines), we see that the vibrational excitation is highly sensitive to the detuning of both frequencies and decays to values of $\langle N_{i} \rangle^{\text{{ss}}} < 50$ already at a detuning of $\Delta=0.1~\text{meV}$. As we showed in Section~\ref{subsec: Directly Solving the Langevin Equation}, the immense vibrational excitation for detunings $\Delta\rightarrow 0$ is caused by the nonconservative part of the adiabatic force. By increasing the detuning, however, the influence of this force becomes less dominant, resulting in the decreasing excitation. We therefore conclude that for the parameters used here, both modes have to be essentially degenerate for the system to exhibit vibrational instabilities.

In the following, we show that when increasing the electronic vibrational coupling, we can observe vibrational instabilities even for detuned modes. In Fig.~\figref{fig: phonon_number_different_detunings}, we show the vibrational excitation of both modes when increasing the electronic-vibrational couplings used in Section~\ref{subsec: Directly Solving the Langevin Equation} by a factor of 5, $\lambda_i\rightarrow5\lambda_i$. Again we observe a immense vibrational excitation for $\Delta\rightarrow0$, clearly indicating a instability of the junction. Even though the vibrational excitation is significantly lower than in the previous case, it is still far above of what would be considered to be a stable junction. Moreover, for this case, we observe that the vibrational excitation is far less sensitive to the detuning of both modes. Although not shown here, the vibrational excitation of both modes below $\Delta=0.3~\text{meV}$ does not significantly change when simulating the dynamics by setting $\mathbf{f}(t)=0$. Therefore, we conclude that the large vibrational excitation below $\Delta=0.3~\text{meV}$ is dominantly caused by current-induced nonconservative forces. Thus, by increasing the electronic-vibrational coupling, we can see vibrational instabilities also for detuned vibrational modes. Note that even the largest detuning displayed in Fig.~\figref{fig: phonon_number_different_detunings} is still a factor of 10 smaller than what has previously been considered in systems exhibiting runaway modes \cite{L2012}.

\section{Conclusion} \label{sec: Conclusion}

In this work, we investigated vibrational instabilities in molecular nanojunctions with multiple vibrational modes that arise from nonconservative current-induced forces as opposed to standard current-induced heating. Although the effect of Joule heating on the stability of such systems is well understood, the mechanisms underlying instabilities from non-conservative forces have only been explored with approximate methods relying on a deterministic eigenmode analysis \cite{Dundas2009,L2010, L2011}. In contrast, our approach used a Markovian Langevin equation combined with NEGFs to calculate the steady-state dynamics of a two-level, two-mode model, from which we calculated the nonequilibrium steady-state electric current and corresponding vibrational excitation. 

Our simulations demonstrated that for the case of degenerate vibrational modes, $\omega_{1} = \omega_{2}$, both modes experience an extremely large vibrational excitation at a much smaller bias voltage than any significant excitation caused by Joule heating, indicating a vibrational instability arising from the nonconservative forces. In contrast, when the frequencies of the vibrational modes were slightly detuned, we observed a significantly lower vibrational excitation of both modes, indicating that the mechanism vanishes for finite detuning. Furthermore, although previous investigations have highlighted the importance of the Berry force in generating instabilities at finite detuning, in our approach, the Berry force had a negligible impact on the vibrational dynamics. 

Next, we applied an eigenmode analysis to the same problem, in which the electronic forces are linearized and the stochastic Langevin equation is reduced to a deterministic equation of motion. Based on the positivity (negativity) of the inverse Q-factor of the corresponding eigenmodes, one can then infer the stability (instability) of the junction. While this approach is consistent with the full Langevin dynamics for degenerate vibrational modes, it is more difficult to interpret for detuned modes, as there are regions of the vibrational coordinate space where the inverse Q-factor is negative and regions where it is positive. Therefore, we concluded that one cannot determine from our eigenmode analysis alone whether the junction is stable or unstable. Finally, we observed that, when increasing the electronic vibrational coupling, the vibrational excitation is far less sensitive to the detuning of both modes. For this case, we found vibrational instabilities also for finite detuning of both vibrational modes.

While the analysis in this paper has been restricted to a model system for which vibrational instabilities have previously been reported, it is straightforward to apply our method to more realistic models of molecular nanojunctions with anharmonic vibrational potentials and nonlinear electronic-vibrational couplings. Moreover, for systems with strong intrasystem interactions, such as electron-electron interactions, one can apply the HEOM and Langevin dynamics approach for an analysis of the vibrational dynamics \cite{Rudge2023,Rudge2024,10.1063/5.0222076}. 

\section*{Acknowledgment}

This work was supported by the Deutsche Forschungsgemeinschaft (DFG) through Research Unit FOR5099, as well as support from the state of Baden-W¨urttemberg through bwHPC and the DFG through Grant No. INST 40/575-1 FUGG (JUSTUS 2 cluster). S.L.R. and R.J.P. thank the Alexander von Humboldt Foundation for support via research fellowships. Moreover, the authors thank Mads Brandbyge for helpful discussions.

\FloatBarrier

\bibliography{references.bib}
\end{document}